\documentclass[]{aa}
\usepackage{graphics,epsfig,txfonts,xspace}
\usepackage{natbib}
\usepackage{longtable}
\usepackage{lscape}
\usepackage{textcomp}

\bibpunct{(}{)}{;}{a}{}{,}

\newcommand{\ergcm}[1]{$\times 10^{#1}$ erg cm$^{-2}$ s$^{-1}$}

\newcommand{\hcm}[1]{$\times 10^{#1}$ cm$^{-2}$}

\newcommand{\nh}{N$_{\rm H}$}

\newcommand{\Halp}{H${\alpha}$}
\newcommand{\ltsima}{$\buildrel < \over \sim$}
\newcommand{\lsim}{\lower.5ex\hbox{\ltsima}}
\newcommand{\gtsima}{$\buildrel > \over \sim$}
\newcommand{\gsim}{\lower.5ex\hbox{\gtsima}}
\newcommand{\msun}{M_{\odot}}
\newcommand{\xmm}{XMM-Newton\xspace}

\def\rahour{\hbox{\ensuremath{^{\rm h}}}}
\def\ramin{\hbox{\ensuremath{^{\rm m}}}}

\newcommand{\xmms}{\hbox{XMMU\,J010147.5-715550}\xspace}

\newcommand{\wdbelmc}{\hbox{XMMU\,J052016.0-692505}\xspace}

\usepackage{color}

\begin{document}
 
\title{A new super-soft X-ray source in the Small Magellanic Cloud:\\ 
       Discovery of the first Be/white dwarf system in the SMC?}
\author{     R.~Sturm\inst{1} 
    \and     F.~Haberl\inst{1}  
    \and     W.~Pietsch\inst{1}  
    \and     M.J. Coe\inst{2} 
    \and     S. Mereghetti\inst{3}  
    \and     N. La Palombara\inst{3} 
    \and     R.~A.~Owen\inst{4} 
    \and     A.~Udalski\inst{5}  
       }

\titlerunning{Discovery of the first Be/WD system in the SMC?}
\authorrunning{Sturm et al.}

\institute{Max-Planck-Institut f\"ur extraterrestrische Physik, Giessenbachstra{\ss}e, 85748 Garching, Germany
	   \and
           School of Physics and Astronomy, University of Southampton, Highfield, Southampton SO17 1BJ, United Kingdom
	   \and
           INAF, Istituto di Astrofisica Spaziale e Fisica Cosmica Milano, via E. Bassini 15, 20133 Milano, Italy
	   \and
           Laboratoire AIM, CEA/DSM - CNRS - Universit\'{e} Paris Diderot, IRFU/SAp, CEA-Saclay, 91191 Gif-sur-Yvette, France
	   \and
           Warsaw University Observatory, Aleje Ujazdowskie 4, 00-478 Warsaw, Poland
	   }

\date{Received 29 July 2011 / Accepted 25 November 2011}

\abstract{The Small Magellanic Cloud (SMC) hosts a large number of Be/X-ray binaries, 
          however no Be/white dwarf system is known so far, although population synthesis calculations predict that they might be more frequent than Be/neutron star systems.} 
         {\xmms was found as a new faint super-soft X-ray source (SSS) with a likely Be star optical counterpart. 
          We investigate the nature of this system and search for further high-absorbed candidates in the SMC.}
	 {We analysed the \xmm X-ray spectrum and light curve, optical photometry, and the $I$-band OGLE III light curve.}
         {The X-ray spectrum is well represented by black-body and white dwarf atmosphere models with highly model-dependent temperature between 20 and 100 eV.
          The likely optical counterpart AzV\,281 showed low near infrared emission during X-ray activity, followed by a brightening in the $I$-band afterwards.
          We find further candidates for high-absorbed SSSs with a blue star as counterpart.}
         {We discuss \xmms as the first candidate for a Be/white dwarf binary system in the SMC.} 

 
\maketitle

\section{Introduction}
\label{sec:introduction}
Super-soft X-ray sources \citep[SSSs, e.g. ][]{2006AdSpR..38.2836K} are a phenomenological class of X-ray sources, 
defined by a very soft X-ray spectrum without significant emission above 1 keV.
They were discovered first in the Magellanic Clouds due to their close distance and low foreground absorption.
The general scenario for a SSS is thermonuclear burning on the surface of an accreting white dwarf (WD), 
which can be stable for certain accretion rates \citep{2007ApJ...663.1269N}. 
SSSs are associated with cataclysmic variables (CVs), symbiotic stars, and post-outburst optical novae.
Super-soft X-ray emission at lower luminosity can e.g. originate from some CVs, cooling neutron stars (NSs), active galactic nuclei (AGN), PG 1159 stars (hot cooling isolated WDs), and planetary nebulae.
Also Be stars \citep[e.g. ][]{2003PASP..115.1153P} are believed to harbour accreting WDs.
By a mechanism not well understood and in combination with high rotation,
Be stars eject material in the equatorial plane causing the build up of a circumstellar decretion disc \citep{2001PASJ...53..119O,2002MmSAI..73.1038Z}.
This disc causes line emission and a near infrared (NIR) excess, 
both showing high variability pointing to the instability of these discs.
How far the generation of the Be phenomenon is related to close binary evolution or another mechanism 
is still under debate \citep{2000ASPC..214..668G}. 
In the case of close binary evolution, matter transfer causes a spin up of the gainer, becoming a Be star, whereas the donor turns into a He star, WD or NS.
Hereafter, accretion from the decretion disc onto the compact object causes X-ray emission.

A high recent star formation \citep{2010ApJ...716L.140A} and low metalicity \citep{2006MNRAS.370.2079D} are 
probably responsible for the remarkably high number of $\sim$90 known Be/X-ray binaries in the SMC.
In these systems a NS accretes matter from the decretion disc of the Be star \citep[for a recent review see ][]{2011Ap&SS.332....1R}.
Binary system evolution models predict, that Be/WD systems might be more frequent than Be/NS systems \citep{1991A&A...241..419P,2001A&A...367..848R}.
\citet{2001A&A...367..848R} obtain an abundance ratio for He star/NS/WD of 2/1/7, but no Be/WD system is known so far in the SMC in contrast to the large number of Be/NS systems.
Similar to an accreting NS, WDs are expected to show hard X-ray emission powered by accretion, 
but at lower luminosity of $10^{29-33}$ erg s$^{-1}$, compared to $10^{34-38}$ erg s$^{-1}$ for NSs \citep{1989A&A...220L...1W}.
Suggested candidates are $\gamma$\,Cas like objects \citep{1995A&A...296..685H,2006A&A...454..265L}.
In addition, nuclear burning on the WD surface can produce super-soft X-ray emission at luminosities of  
$10^{35-38}$ erg s$^{-1}$, which might be absorbed by the decretion disc in most cases, 
complicating the discovery of such systems \citep{1991A&A...248..139A}.
Only one Be/WD SSS system has been proposed to date, \wdbelmc \citep{2006A&A...458..285K} in the Large Magellanic Cloud (LMC).
To investigate the observational boundary conditions for Be/WD systems and to constrain the evolutionary models in general,
it is of high importance to find more such systems.

The \xmm \citep{2001A&A...365L...1J} large programme survey of the SMC (Haberl et al. 2012, in preparation), in combination with archival observations, 
provides a flux limited sample of X-ray sources of the central field of the SMC.
This enables comprehensive studies of SSSs \citep[e.g.][]{2010A&A...519A..42M,2011A&A...529A.152S} 
as well as Be/X-ray binaries \citep[e.g.][]{2011A&A...527A.131S,2011MNRAS.414.3281C}.
In a search for new faint SSSs in our point source catalogue (Sturm et al. 2012, in preparation), we found one candidate, correlating with an early type SMC star.
In this study, we present the results of the analysis of the \xmm EPIC X-ray data, together with optical information.
This led to the discovery of the first candidate Be/WD system in the SMC -- \xmms\ -- and 
illustrates, that \wdbelmc in the LMC is not a unique case.

\section{Analyses of X-ray data and results}

XMMU\,J010147.5-715550 was found in a search for SSS candidates in the combined SMC survey and archival data.
The data were collected with EPIC-pn \citep{2001A&A...365L..18S} and EPIC-MOS \citep{2001A&A...365L..27T}.
The source was detected three times in images of calibration observations of the supernova remnant 1E0102-72 
at off-axis angles of $\sim$10.5\arcmin\ on MJD 51650, 51651, and 52014.
The error weighted average of the astrometric boresight corrected best-fit positions is RA (J2000)=01\rahour01\ramin47\fs58 
and Dec (J2000)=$-$71\degr55\arcmin50\farcs7 with a $1\sigma$ uncertainty of 0.85\arcsec.

There is no star within 10\arcsec, sufficiently bright to cause optical loading which might fake a SSS.
The multiple detections rule out detector effects for spurious detections.
Thus we conclude the veritableness of the X-ray source.
The source is also listed in the \xmm incremental source catalogue \citep[2XMMi\,J010147.5-715550, ][]{2009A&A...493..339W}.

Detected fluxes $F$, detection likelihoods $ML$ and net exposures of the individual observations are summarised in Table~\ref{tab:obs}.
In later calibration observations, the source was marginally detected on MJD 52269, 52385, and 52412
and not detected above a detection likelihood $ML>2$ in any of the 21 observations after this date. In the later observations, the source was mostly not covered by EPIC-pn.
We merged the EPIC-MOS1 and -MOS2 data of these latter observations and measured counts in circular source and background regions with 20\arcsec\ radius.
We obtain 18$\pm$16 and 23$\pm$18 net counts in the (0.2--1.0) keV band.
For vignetting corrected exposures of 182 ks from MOS1 and 235 ks from MOS2,
this translates into a 3$\sigma$ flux limit of 10.2 \ergcm{-16} 
and indicates a significant source variability of a factor of $\sim$10.

\begin{table}
\caption[]{X-ray observations of \xmms.}
\begin{center}
\begin{tabular}{lccrr}
\hline\hline\noalign{\smallskip}
\multicolumn{1}{c}{ObsID} &
\multicolumn{1}{c}{MJD} &
\multicolumn{1}{c}{$F$ in (0.2--1.0) keV  } &
\multicolumn{1}{c}{$ML$}  &
\multicolumn{1}{c}{Exp.}  \\
\multicolumn{1}{l}{} &
\multicolumn{1}{c}{} &
\multicolumn{1}{c}{($10^{-16}$ erg cm$^{-2}$ s$^{-1}$)} &
\multicolumn{1}{c}{}  &
\multicolumn{1}{c}{(ks)}  \\
\noalign{\smallskip}\hline\noalign{\smallskip}
 0123110201              &     51650.9 & 92.8 $\pm$ 11.9      &              88.0 &        17.9 \\  
 0123110301              &     51651.3 & 74.3 $\pm$ 13.7      &              31.5 &        11.7 \\  
 0135720601              &     52014.1 & 99.3 $\pm$ 16.6      &              36.0 &        13.4 \\  
 0135720801$^{\dagger}$ &     52269.0 & 32.0 $\pm$ 10.5      &               5.8 &        27.6 \\  
 0135720901              &     52385.0 & 25.1 $\pm$ 11.1      &               6.6 &        10.1 \\  
 0135721001              &     52412.5 & 12.4 $\pm$ 8.8       &               2.1 &         9.4 \\  
 0135721001$^{\dagger}$ &     52412.7 & 66.1 $\pm$ 29.8      &               2.9 &         7.5 \\  
 0135721101$^{\dagger}$ &     52560.3 & 15.1 $\pm$ 11.8      &               0.0 &        23.1 \\  
 0135721301$^{\dagger}$ &     52622.3 &  $<$ 11.1            &               0.2 &        28.4 \\  
 0135721401$^{\dagger}$ &     52749.8 &  $<$ 12.4            &               0.1 &        30.4 \\  
 0135721501              &     52939.5 &  $<$ 9.7             &               0.1 &        24.2 \\  
 0135721701              &     52959.4 &  $<$ 7.4             &               0.0 &        25.2 \\  
 0135721901$^{\dagger}$ &     53123.5 &  $<$ 12.3            &               0.2 &        33.0 \\  
 0135722001$^{\dagger}$ &     53304.5 &  $<$ 11.2            &               1.4 &        31.6 \\  
 0135722101              &     53316.3 & 8.6 $\pm$ 5.0        &               0.6 &        24.0 \\  
 0135722301$^{\dagger}$ &     53317.1 &  $<$ 26.5            &               0.0 &        14.0 \\  
 0135722401$^{\dagger}$ &     53292.6 &  $<$ 20.8            &               0.0 &        30.4 \\  
 0135722501$^{\ddag}$     &     53478.1 &  $<$ 15.0            &               0.0 &        23.8 \\  
 0135722601$^{\dagger}$ &     53679.5 &  $<$ 7.9             &               0.0 &        29.0 \\  
 0135722701$^{\ddag}$     &     53845.3 & 17.5 $\pm$ 10.9      &               1.3 &        30.2 \\  
 0412980101$^{\dagger}$ &     54044.2 &  $<$ 9.2             &               0.1 &        31.0 \\  
 0412980201$^{\ddag}$     &     54215.7 &  $<$ 11.8            &               0.0 &        19.6 \\  
 0412980301$^{\dagger}$ &     54399.6 & 12.0 $\pm$ 6.7       &               0.1 &        35.0 \\  
 0412980501$^{\ddag}$     &     54575.6 &  $<$ 16.7            &               0.0 &        22.0 \\  
 0412980701$^{\dagger}$ &     54785.0 &  $<$ 13.9            &               0.0 &        28.6 \\  
 0412980801$^{\ddag}$     &     54934.2 &  $<$ 15.3            &               0.0 &         6.5 \\  
 0412980901$^{\dagger}$ &     55125.5 & 15.4 $\pm$ 10.0      &               0.5 &        28.6 \\  
 0412981001$^{\ddag} $    &     55307.2 & 27.2 $\pm$ 16.6      &               0.2 &        27.3 \\  
\noalign{\smallskip}\hline
\end{tabular}
\end{center}
\tablefoot{
Combined EPIC-pn and EPIC-MOS source detection.
Fluxes are calculated from count rates by assuming the best-fit black-body model. 
 Uncertainties are for 1$\sigma$.
 For observations marked with $\dagger$, the source was not covered by EPIC-pn, but both MOS cameras.
 For observations, marked with $\ddag$, the source was in the FoV of EPIC-MOS2 only. 
 In all other observations, the source was observed with all EPIC instruments.
 }
\label{tab:obs}
\end{table}

Hardness ratios can be used to characterise the source spectrum. 
They are defined as $HR_1=(R_2-R_1)/(R_2+R_1)$ and $HR_2=(R_3-R_2)/(R_3+R_2)$, 
where $R_1$, $R_2$, and $R_3$ are the count rates in the three standard energy bands (0.2--0.5) keV, (0.5--1.0) keV, and (1.0--2.0) keV, respectively.
For the first three observations, we obtain 
$HR_1= -0.10\pm0.12$, $-0.69\pm0.15$, and $-0.13\pm0.16$ and 
$HR_2= -0.85\pm0.11$, $-0.09\pm0.42$, and $-1.00\pm0.13$.
Based on hardness ratio criteria \citep[see e.g. ][]{2004A&A...426...11P}, the source was classified as SSS in our analysis.
The 2XMMi catalogue gives $HR_1=-0.45\pm0.07$ and $HR_2=-0.93\pm0.07$, consistent with our classification.

We extracted EPIC-pn source and background spectra from the first three observations, 
using the XMM-Newton SAS version 10.0.0\footnote{Science Analysis Software (SAS), http://xmm.esac.esa.int/sas/}.
We selected single- and double-pixel events with {\tt FLAG=0} from a source region, optimized in size for signal to noise ratio by {\tt eregionanalyse}, 
and a background region in a point source free area.
The spectra have 55, 27, and 15 net counts in the (0.15--2.0) keV band.
We fitted black-body (bb), power law (pl), thermal plasma emission (apec), and several non local thermal equilibrium \citep[nlte, ][]{2003ASPC..288..103R}
models using  {\tt xspec} \citep{1996ASPC..101...17A} version 12.6.0k with C statistic \citep{1979ApJ...228..939C}.
The photo-electric absorption is described by a Galactic component with column density fixed to a maximum of N$_{\rm H}^{\rm gal}$ = 6\hcm{20} \citep{1990ARA&A..28..215D} 
and elemental abundances according to \citet{2000ApJ...542..914W} and an additional (SMC + source intrinsic) absorption with free column density and 
abundances set to 0.2 for elements heavier than helium \citep{1992ApJ...384..508R}. 
To test for a Be/X-ray binary with soft excess, we fitted the combination of a black-body and power-law model (bb+pl).

\begin{figure}
  \resizebox{\hsize}{!}{\includegraphics[angle=-90,clip=]{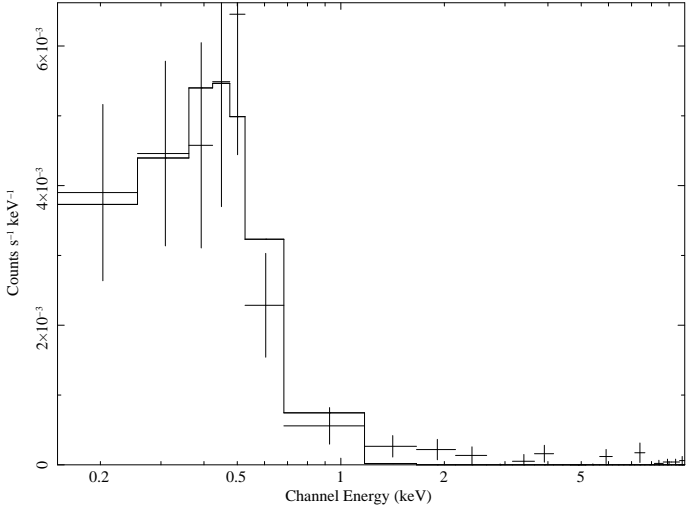}}

  \caption{
    Co-added spectrum of three EPIC-pn observations of \xmms together with the best-fit black-body model.
  }
  \label{fig:spectrum}
\end{figure}

Fitting the three EPIC-pn spectra simultaneously by assuming a common black-body temperature, we found no significant differences for the fluxes
(relative: 1.0, 0.91$^{+0.50}_{-0.36}$, and 0.93$^{+0.61}_{-0.41}$).  Thus, we fitted the same models to all three spectra.
The co-added spectrum, rebinned to 3$\sigma$ significance per bin for better illustration, together with the best-fit bb model is shown in Fig.~\ref{fig:spectrum}.
The detected flux in the (0.2--1.0) keV band is $(9.4\pm2.4) \times 10^{-15}$ erg cm$^{-2}$ s$^{-1}$.
The best-fit model parameters are listed in Table~\ref{tab:spec}. 
Statistical uncertainties and limits are for 90\% confidence.
To estimate the goodness of the fit, in addition to the C statistics we give also the $\chi^2$ statistics, which is derived by fitting the models to spectra, binned to a minimum of 10 cts/bin.
These fits result in similar parameters as those given in Table~\ref{tab:spec} derived with C statistics.
Most models give an acceptable fit. 
Possible physical scenarios for these models are discussed in Sec.~\ref{sec:discussion}.
To demonstrate the parameter dependence on modelling,
we give the best-fit values for two nlte models for surface gravity $\log(g)=9$ with solar (nlte$_{\rm S}$, $z=0$) and halo (nlte$_{\rm H}$, $z=-1$) abundances.

\begin{table}[ht!]
\caption[]{Spectral fitting results for several models.}
\begin{center}
\begin{tabular}{lcccccrr}
\hline\hline\noalign{\smallskip}
\multicolumn{1}{c}{model} &
\multicolumn{1}{c}{N$_{\rm H}^{\rm gal}$} &
\multicolumn{1}{c}{N$_{\rm H}^{\rm smc}$} &
\multicolumn{1}{c}{$kT$} &
\multicolumn{1}{c}{$\Gamma$} &
\multicolumn{1}{c}{$F$} &
\multicolumn{1}{c}{C} &
\multicolumn{1}{c}{$\chi_{\rm red}^2$} \\
\noalign{\smallskip}\hline\noalign{\smallskip}
bb             &   6$^{(f)}$       &   $<$9         &    $97_{-21}^{+15}$       &   --              &   22        &   90.6     & 0.95 \\ \vspace{0.5mm} 
nlte$_{\rm S}$         &   6$^{(f)}$       &  $<$3          &    $80_{-6}^{+3}$        &   --              &   22                  &   91.2     & 1.02 \\  \vspace{0.5mm}
nlte$_{\rm H}$         &   6$^{(f)}$       &  $53_{-6}^{+7}$&    $26_{-1}^{+2}$        &   --              &   170000  &   91.7     & 0.74 \\  \vspace{0.5mm} 
apec           &   $<$46           &   --             &    $92_{-65}^{+97}$     &   --              &   35        &  107.9     & 1.72 \\  \vspace{0.5mm}
pl             &   6$^{(f)}$       &   $<$15         &          --            & $5.0_{-1.2}^{+2.3}$  &   120       &   89.5     & 0.97 \\  \vspace{0.5mm}
bb+pl          &   6$^{(f)}$       &   $<$14         &    $88_{-24}^{+17}$       &  1.0$^{(f)}$       &   23+8      &  86.6      & 0.95 \\  
\noalign{\smallskip}\hline
\end{tabular}
\end{center}
\tablefoot{N$_{\rm H}$ in $10^{20}$ cm$^{-2}$, $kT$ in eV, unabsorbed flux $F$ in the (0.2--1.0) keV band in $10^{-15}$ erg cm$^{-2}$ s$^{-1}$. Fixed values are marked with $^{(f)}$.}
\label{tab:spec}
\end{table}

\section{The optical counterpart}

The  Magellanic Cloud Photometric Survey \citep[MCPS, ][]{2002AJ....123..855Z} lists two possible counterparts within the X-ray position uncertainty.
The fainter one ($V=17.0$ mag), with angular separation of 2.1\arcsec, has a proper motion of $\sim$580 mas yr$^{-1}$ in the USNO-B1.0 catalogue \citep{2003AJ....125..984M}.
It is therefore a Galactic foreground star.

\begin{figure}
  \resizebox{0.9\hsize}{!}{\includegraphics[angle=0,clip=]{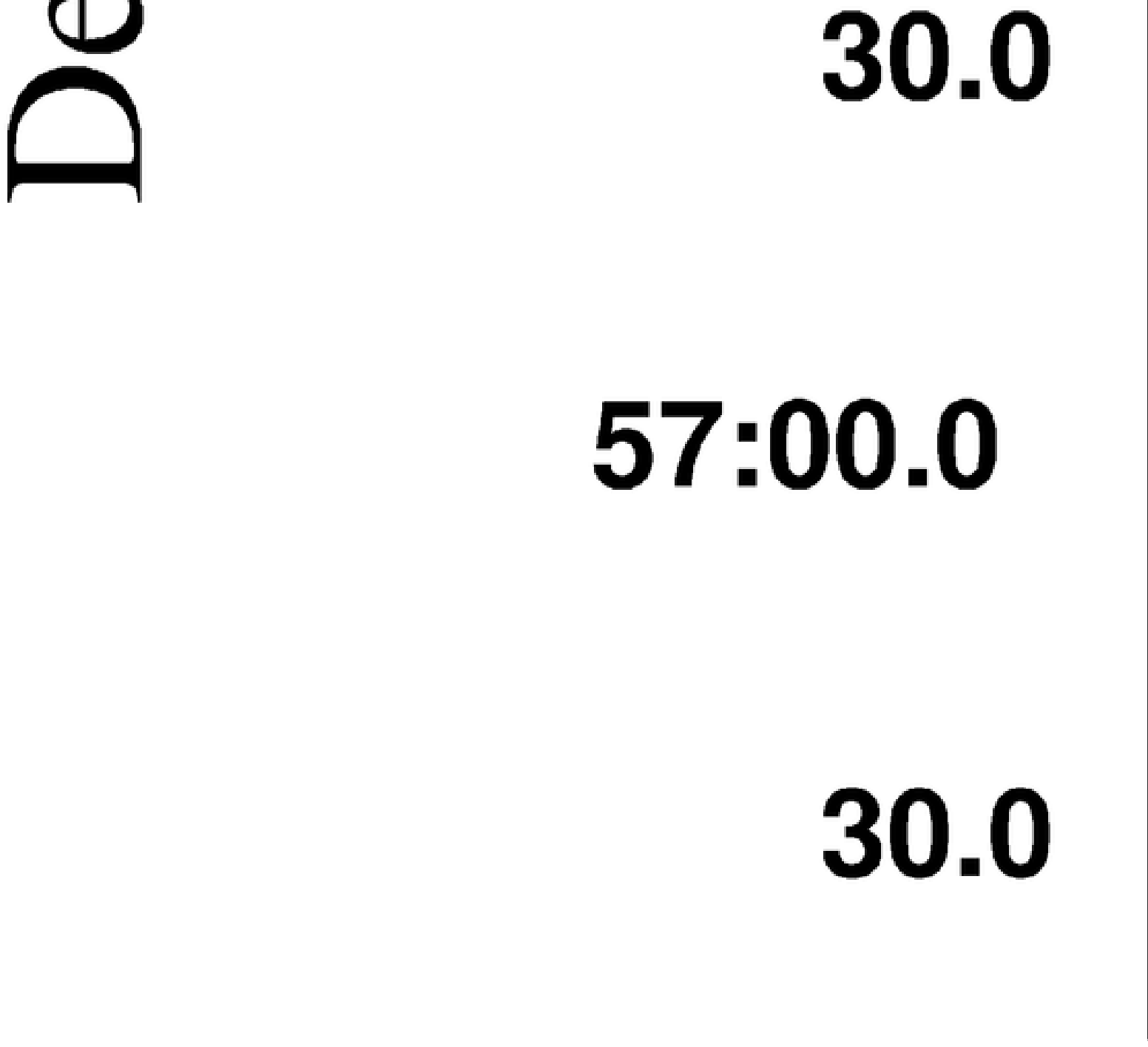}}
  \caption{
    $I$-band finding chart from OGLE. Lines mark OGLE-III SMC113.5 8.
    The zoom-in shows the optical counterpart over plotted with the positions of the 3 X-ray detections with 1$\sigma$ radii.
  }
  \label{fig:fc}
\end{figure}

The brighter counterpart at a distance of 1.4\arcsec, has 
$U=(13.26\pm0.03)$ mag, 
$B=(14.40\pm0.02)$ mag, 
$V=(14.47\pm0.04)$ mag, and 
$I=(14.30\pm0.04)$ mag.
This star (\object{AzV\,281}) was classified as a B0 star in the SMC by \citet[][]{1975A&AS...22..285A} and as O7III by \citet{2002ApJS..141...81M}.
The star corresponds to the \Halp\ emission line object \object{[MA93] 1284} and \object{LIN 384} \citep{1993A&AS..102..451M,1961AJ.....66..169L}.
Therefore, the optical counterpart most likely is an O7IIIe--B0Ie star.
An $I$-band image from OGLE III \citep[][]{2008AcA....58...69U} of the surroundings of this star (OGLE III SMC113.5 8) is shown in Fig.~\ref{fig:fc}.

\begin{figure}
  \resizebox{\hsize}{!}{\includegraphics[angle=0,clip=]{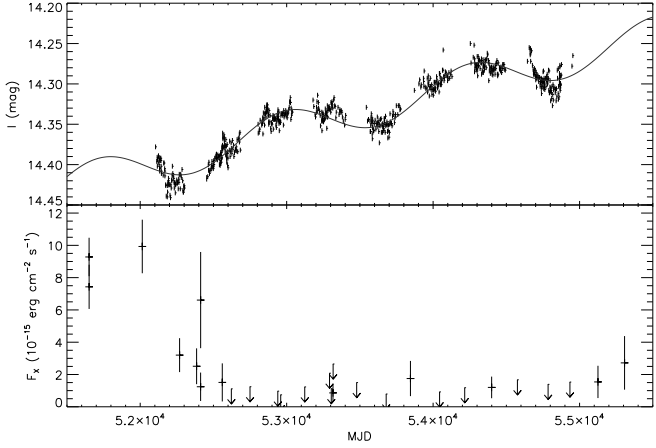}}

  \caption{
    {\it Top:} $I$-band light curve from OGLE III.
    {\it Bottom:} X-ray flux in the (0.2--1.0) keV band. 
  }
  \label{fig:lc}
\end{figure}

The $I$-band OGLE III light curve of AzV\,281 is compared to the X-ray light curve in Fig.~\ref{fig:lc}. 
It exhibits an increasing trend in brightness superimposed by sinusoidal variations.
A fit of a linear plus sine function (solid line in Fig.~\ref{fig:lc}) results in a constant flux increase of (-0.01687$\pm$0.00005) mag yr$^{-1}$,
a sine amplitude of (0.0240$\pm$0.0002) mag, and a period of (1264$\pm$2) days.
Uncertainties are given for 1$\sigma$ confidence.
A Lomb Scargle periodogram did not reveal any further significant variability from the source.

\begin{figure}
  \resizebox{\hsize}{!}{\includegraphics[clip=]{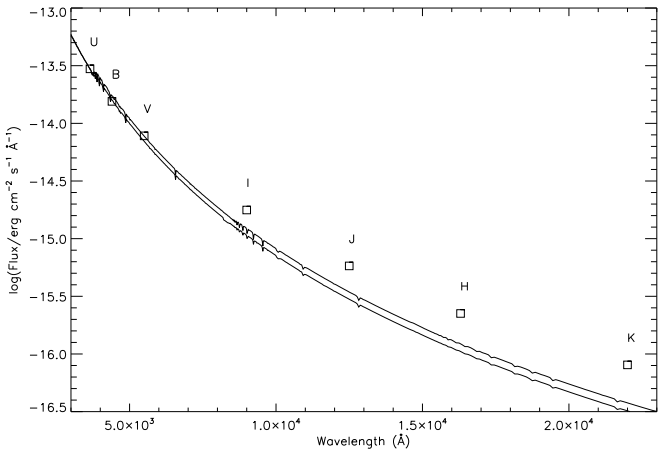}}
  \caption{
   Dereddened optical \citep{2002AJ....123..855Z} and NIR \citep{2007PASJ...59..615K} fluxes (squares) compared to a Kurucz model atmosphere for a B1V (upper line) and an O7V (lower line) star both normalised to the $U$-band flux.
  }
  \label{fig:spec_opt}
\end{figure}

The star is included in the near infrared (NIR) 
2MASS catalogue \citep[][ID 01014789-7155512, 
$J = (14.24\pm0.03)$ mag, 
$H = (14.21\pm0.03)$ mag, 
$K = (14.04\pm0.05)$ mag,
on MJD 51034]{2006AJ....131.1163S}
and the IRSF/SIRIUS catalogue \citep[][ID 01014790-7155512, 
$J = (14.44\pm0.02)$ mag, 
$H = (14.36\pm0.01)$ mag, 
$K = (14.24\pm0.03)$ mag,
on MJD 52535, magnitudes are converted to the 2MASS system]{2007PASJ...59..615K}. 
In Fig.~\ref{fig:spec_opt}, the optical and NIR fluxes, corrected for $E(B-V)=0.08$, 
are compared to Kurucz stellar atmosphere models for a B1V and an O7V star both normalised to the dereddened $U$-band flux.
The NIR excess and NIR variability 
provide further evidence for a Be star.
Remarkably, the NIR flux decreased significantly during or before the time of the X-ray detections.

\begin{table*}[ht!]
\caption[]{Absorbed SSS candidates with possible early type star counterpart in the SMC.}
\begin{center}
\begin{tabular}{lccccccccrrrcc}
\hline\hline\noalign{\smallskip}
\multicolumn{6}{c|}{X-ray} &
\multicolumn{5}{c|}{MCPS} &
\multicolumn{2}{c|}{MA93} &
\multicolumn{1}{c}{} \\
\multicolumn{1}{c}{RA} &
\multicolumn{1}{c}{Dec} &
\multicolumn{1}{c}{$\Delta r$} &
\multicolumn{1}{c}{$HR_1$} & 
\multicolumn{1}{c}{$HR_2$}  &
\multicolumn{1}{c|}{$ML$}  &
\multicolumn{1}{c}{$d$}  &
\multicolumn{1}{c}{$U$}  &
\multicolumn{1}{c}{$B$}  &
\multicolumn{1}{c}{$V$}  &
\multicolumn{1}{c|}{$I$}  &
\multicolumn{1}{c}{$d$}  &
\multicolumn{1}{c|}{cat. No}  &
\multicolumn{1}{c}{comment} \\
\multicolumn{1}{c}{(J2000)} &
\multicolumn{1}{c}{(J2000)} &
\multicolumn{1}{c}{(\arcsec)} &
\multicolumn{1}{c}{} & 
\multicolumn{1}{c}{}  &
\multicolumn{1}{c|}{}  &
\multicolumn{1}{c}{(\arcsec)}  &
\multicolumn{1}{c}{(mag)}  &
\multicolumn{1}{c}{(mag)}  &
\multicolumn{1}{c}{(mag)}  &
\multicolumn{1}{c|}{(mag)}  &
\multicolumn{1}{c}{(\arcsec)}  &
\multicolumn{1}{c|}{[MA93]}  &
\multicolumn{1}{c}{} \\
\noalign{\smallskip}\hline\noalign{\smallskip}
  01 17 40.44   &  -73 30 50.7  &  1.3  &  0.6$\pm$0.2  &  -0.3$\pm$0.2  &  11.8  &  1.3  &  13.3  &  14.1  &  14.2  &  14.1  &  0.4  &  1845 & HMXB  \\    
  01 05 16.05   &  -72 00 25.0  &  2.1  &  0.9$\pm$0.2  &  -1.0$\pm$0.1  &  10.6  &  5.2  &  14.9  &  15.8  &  16.0  &  16.2  &  --  &  -- & --  \\         
  00 59 54.57   &  -72 15 30.5  &  1.8  &  1.0$\pm$0.2  &  -0.7$\pm$0.2  &  12.3  &  4.4  &  15.8  &  16.8  &  16.9  &  17.0  &  --  &  -- & --  \\         
  00 49 13.54   &  -73 25 60.0  &  1.8  &  0.3$\pm$0.2  &  -0.7$\pm$0.2  &  15.5  &  0.8  &  14.6  &  15.4  &  15.4  &  15.1  &  --  &  -- & --  \\         
  01 01 05.01   &  -72 04 43.5  &  3.3  &  0.3$\pm$0.2  &  -0.7$\pm$0.2  &  23.7  &  5.5  &  15.3  &  16.2  &  16.4  &  16.6  &  --  &  -- & Gal. star?  \\ 
  00 53 55.61   &  -72 26 44.4  &  2.1  &  0.7$\pm$0.3  &  -0.5$\pm$0.2  &  10.0  &  1.3  &  13.6  &  14.6  &  14.7  &  14.6  &  --  &  -- & HMXB  \\       
  00 51 52.50   &  -72 31 47.8  &  1.6  &  0.5$\pm$0.2  &  -0.5$\pm$0.2  &  10.8  &  3.1  &  13.7  &  14.8  &  14.4  &  14.4  &  3.4  &  506 & Gal. star?\\ 
  00 55 07.25   &  -72 08 25.7  &  1.7  &  -0.1$\pm$0.3  &  -0.6$\pm$0.3  &  8.6  &  3.9  &  16.0  &  16.8  &  16.9  &  16.9  &  --  &  -- & HMXB?  \\      
  00 55 08.45   &  -71 58 26.7  &  1.4  &  -0.2$\pm$0.2  &  -0.9$\pm$0.4  &  14.0  &  4.6  &  14.2  &  15.2  &  15.4  &  15.6  &  --  &  -- & --  \\        
  00 54 14.97   &  -72 14 43.1  &  3.4  &  1.0$\pm$0.5  &  -1.0$\pm$0.5  &  11.5  &  5.3  &  15.8  &  16.5  &  16.6  &  16.8  &  12.0  &  744 & --  \\      
\noalign{\smallskip}\hline
\end{tabular}
\end{center}
\tablefoot{$\Delta r$: 1$\sigma$ X-ray position uncertainty, including the systematic error. 
           $HR$: X-ray hardness ratios.
           $ML$: detection likelihood.
           $d$: Angular distance to the counterpart from MCPS \citep{2002AJ....123..855Z} and MA93 \citep{1993A&AS..102..451M}.
          }
\label{tab:cand}
\end{table*}

\section{Discussion}
\label{sec:discussion}

We analysed \xmm X-ray data and OGLE III photometry of \xmms.
In the following we will discuss possible classifications of the source:

{\it A Galactic star} 
 can produce soft coronal X-ray emission. In this case the fainter object could be the true counterpart.
 But the emission of stars extends to higher energies than we see from our SSS causing a positive $HR_1$ in general \citep{2004A&A...426...11P,2011A&A...534A..55S}.
 Also, the derived $\chi_{\rm red}^2$ of the apec model suggests formally that a description of the spectrum by a thermal plasma is less likely than by the other models.

{\it A cooling isolated compact object}, i.e. a NS or PG 1159 star,
 can emit super-soft X-rays.  We would not expect to see an optical counterpart.
 In both cases, we cannot explain the variability, seen in the X-ray light curve.

For an {\it AGN} 
 with very soft X-ray emission and with a faint undetected optical counterpart, we expect to see the Galactic and total SMC absorption in the line of sight.
 The total SMC column density in the direction of the source is $N_{\rm H, SMC} = 5.1 \times 10^{21}$ cm$^{-2}$ \citep{1999MNRAS.302..417S}.
 We derive a  $N_{\rm H, SMC} < 1.5 \times 10^{21}$ cm$^{-2}$ for the power-law model, which seems to contradict the AGN assumption.

{\it A CV in the SMC} 
 which by chance correlates with AzV\,281 can account for the super-soft emission. 
 However, there is no accretion powered system known with super-soft emission exceeding $10^{33}$ erg s$^{-1}$, 
 which is  below the observed luminosity of \xmms for the distance of the SMC.
 In the case of nuclear surface burning, higher luminosities can be reached (e.g. CAL 83 in the LMC).
 This would need a highly obscured WD, as discussed below.
 Another possibility for variable super-soft X-ray emission would be the rare case of a nova explosion. Here,
 super-soft emission is observed up to 2--5 years \citep{2011A&A...533A..52H} after outburst.
 However, a nova explosion in this system is not known, which of course could have been missed.
 Due to the low nova rate in the SMC (only 3 novae were discovered during the last 10 years) this is an unlikely scenario.
 
{\it A Galactic CV} 
 cannot be excluded with the available data. E.g. the Galactic star could contain a WD companion.
 In this case, a soft intermediate polar \citep{1995A&A...297L..37H} could account for the spectral properties,
 but only a handful of these systems is known in the Galaxy so far.

 During the \xmm SMC survey data analysis, we did not find any other SSS candidate correlating with an early-type emission line star.
 To estimate the chance for a random coincidence of a SSS or SSS candidate (14 in total) with a blue emission line star, 
 we used a subset of early type SMC stars ($V<17$ mag, $-0.5$ mag $<B-V< 0.5$ mag, $-1.5$ mag $<U-B< -0.2$ mag) from the MCPS
 which correlate with an emission line object from MA93.
 This sample (1748 sources inside the observed \xmm\ field) was correlated with X-ray sources, which fulfill our criteria for super-soft X-ray emission 
 and whose coordinates were shifted by multiples of the maximal correlation distance.
 We accepted correlations with an angular separation of $d\leq3.439\times (\Delta r_{1}^2 + \Delta r_{2}^2 )^{1/2}$,
 where $\Delta r$ was estimated to 0.3\arcsec and 2.0\arcsec for the MCPS and MA93 sources, respectively. 
 The \xmm\ sources have source-individual position uncertainties of $\sim$1\arcsec.
 In 440 runs, we obtained only 3 chance correlations.
 Analogous, using a set of USNO-B1.0 sources with proper motions $>$50 mas yr$^{-1}$ (134780 sources) results in 773 chance coincidences.
 This further affirms AzV\,281 as the most likely optical counterpart.
 Also the coincidence of the X-ray emission with the NIR flux minimum further supports the identification of the super-soft X-ray source with AzV\,281.

From a {\it Be/NS system} we would expect outbursts with hard X-ray emission during accretion, 
 which has not been observed for this system, although it was monitored regularly during the last 11 years.
 Also, observing only supersoft X-ray emission is atypical for a Be/NS system, as dominant hard emission is expected in additon to a soft excess \citep{2004ApJ...614..881H}.
 Even in the extreme case of an outburst of RX\,J0103.6-7201 \citep{2008A&A...491..841E}, a hard X-ray component is present and the soft component extends to higher energies.
 From the bb+po model we obtain a limit for $L_{\rm bb}/L_{\rm X}$ of $>$0.5 in the (0.15-10.0) keV band, proofing the dominance of the thermal component.
 Therefore, we argue, that the system is less likely a classical Be/X-ray binary.

In contrast to the former possibilities, a {\it WD/Be system} can account for both the super-soft X-ray emission and the variability. 
  Black-body emission is a crude approximation to the spectra of SSSs \citep{2010AN....331..146R}.
  Physically more meaningful are nlte models, but higher statistics and high resolution spectra would be necessary to constrain the parameters.
  To demonstrate the parameter dependence on elemental abundances we list two nlte models in Table~\ref{tab:spec}.
  Solar abundances result in similar parameters as the black-body model with an emission radius of (73--106) km compared to (19--75) km from the bb model. 
  Halo abundances result in a higher absorption, higher luminosity and an emission radius of (90000--190000) km.
  For an object in the SMC we expect abundances between solar and halo values \citep{1992ApJ...384..508R,2010A&A...520A..85D}.
  Therefore the black-body emission radius is rather a lower limit. 
  This further demonstrates, that luminosity and absorption can not be determined uniquely for \xmms
  and that the spectra might be compatible with absorbed emission from a WD and luminosities sufficiently high for surface burning.

The presumable optical counterpart shows properties typical for a Be star.
This makes the system the second candidate for a Be/WD binary after \wdbelmc in the LMC \citep{2006A&A...458..285K}.

The X-ray turn-off around MJD 52500 can be caused by an exhaustion of the nuclear burning on the WD, 
as it occurs for post-nova SSSs, 
if the accretion rate is too low for stable nuclear burning.
In the case of ongoing mass transfer, reignition of nuclear burning on the WD is possible in a nova explosion, followed by another super-soft X-ray emission state.
More likely, the decrease in soft X-rays in the later observations might be explained by increasing photo-electric absorption 
due to a build-up of the decretion disc around the Be star.
An additional absorbing column density of \nh$\gtrsim 10^{22}$ cm$^{-2}$ would be required for a non-detection of \xmms\ in the merged images.
Be stars can loose and rebuild their discs on a time scale of some years including intervals without a disc of some hundred days \citep{2011Ap&SS.332....1R,2010ApJ...709.1306W}.
The decrease of the NIR flux before or during the X-ray active phase 
and the low photo-electric absorption during the three X-ray detections suggest this scenario.
The X-ray detections have only a small overlap with the optical light curve, 
but since X-ray turn-off, the $I$-band shows an increase in flux,
which can be caused by the recreation of a disc.
We note, that the optical light curve of \wdbelmc \citep[cf. Fig. 4 in ][]{2006A&A...458..285K} 
has similar characteristics of both a long-term increase in the $I$-band and a periodicity around one thousand days.

The low X-ray absorption suggests a location of the system on the near side of the SMC bar.
Also, the system intrinsic absorption along the line of sight must have been low during the X-ray detections.
If the 1264 d period is caused by binarity, we get a relatively large separation of 5.7 AU ($\sim$$140 R_{\ast}$) for $M_{\rm WD}=1\msun$ and $M_{\rm \ast}=15\msun$.
But we note, that the realness and nature of the periodicity are uncertain, since the OGLE light curve covers only about two cycles.
While \citet{2001A&A...367..848R} argue that most Be/WD systems are not detectable in X-rays due to absorption by the decretion disc,
truncation of the disc, as it is known from Be/NS systems with low excentricity \citep{2001A&A...377..161O} might only allow accretion during disc instabilities.
Also, the absorption by the SMC has to be low, to be able  to observe such systems in X-rays. 
We estimate the completeness in our SMC catalogue in the (0.2--1.0) keV band to a flux limit of 2.5\ergcm{-15}, 
which would require a SMC absorption of less than 3 $\times 10^{21}$ cm$^{-2}$ for the best-fit black-body model.
Therefore, \xmms might be the tip of the iceberg of the SMC Be/WD system population.

\section{Other Be/WD candidates in the SMC}

In the SMC, we did not find  any other SSS candidate correlating with an early-type star.
However, the search criteria were optimised to find standard SSS with low absorption.
We studied the possibility of intermediate and strongly absorbed but more luminous Be/WD systems.
Assuming the black-body model as above, we would expect hardness ratios of $HR_1=0.81$ and $HR_2=-0.68$ in the absorbed case (N$_{\rm H}^{\rm smc}=10^{22}$ cm$^{-2}$).
We found 10 additional candidates, listed in Table~\ref{tab:cand}, with $HR_2+\Delta HR_2<0$ and with an angular separation $d\leq3.439\times (\Delta r_{\rm X}^2 + \Delta r_{\rm opt}^2 )^{1/2}$  
to an early type SMC star ($V<17$ mag, $-0.5$ mag $<B-V< 0.5$ mag, $-1.5$ mag $<U-B< -0.2$ mag) from the MCPS.
Three of the counterparts are emission line objects in the list of \citet{1993A&AS..102..451M}.
Using the same method as above, we estimate $\sim$4.3 coincidences by chance (1884 correlations in 440 runs).
Since also stars fall into the X-ray selection criteria, the number of chance coincidences increases.

We note, that three sources correlate with known Be/NS systems, which apparently were detected in a relatively soft low-luminosity state. 
Additional investigation of this sample is necessary.

\section{Summary}
We report the discovery and analysis of a new SSS in the SMC, having a presumable early type star companion similar to \wdbelmc in the LMC.
This star has a variable NIR excess and emission lines, typical for a Be star.
This makes the system the first candidate for a Be/WD system in the SMC. 
The X-ray turn off of the source around MJD 52400 can be caused by an increasing photo-electric absorption, 
possibly caused by the build up of a decretion disc around the Be star.
We list further candidates for brighter but more absorbed Be/WD systems.

\begin{acknowledgements}
This publication is based on observations with XMM-Newton, an ESA Science Mission with instruments and contributions 
directly funded by ESA Member states and the USA (NASA).
The XMM-Newton project is supported by the Bundesministerium f\"ur Wirtschaft und 
Technologie/Deutsches Zentrum f\"ur Luft- und Raumfahrt (BMWI/DLR, FKZ 50 OX 0001)
and the Max-Planck Society.
The OGLE project has received funding from the European Research Council
under the European Community's Seventh Framework Programme
(FP7/2007-2013) / ERC grant agreement no. 246678 to AU.
RS acknowledges support from the BMWI/DLR grant FKZ 50 OR 0907.
SM and NLP thanks the support of the agreement ASI-INAF I/009/10/0.
\end{acknowledgements}

\bibliographystyle{aa}
\bibliography{../general,../auto}

\end{document}